# Controlling liquid-liquid phase behavior with an active fluid

Alexandra M. Tayar[a], Fernando Caballero[a], Trevor Anderberg[a], Omar A. Saleh[b,a,c], M. Cristina Marchetti[a,c], Zvonimir Dogic[a,c].

[a]Department of Physics, University of California, Santa Barbara, CA 93106.
[b]Materials Department, University of California, Santa Barbara, California 93106, USA.
[c]Biomolecular Science and Engineering Program, University of California, Santa Barbara, California 93106, USA.

**Abstract:** Demixing of binary liquids is a ubiquitous transition, which is explained using a well-established thermodynamic formalism that requires equality of intensive thermodynamics parameters across the phase boundaries. Demixing transitions also occur when binary fluid mixtures are driven away from equilibrium, for example, by external shear flow. Predicting demixing transition under non-equilibrium non-potential conditions remains, however, a challenge. We drive liquid-liquid phase separation of attractive DNA nanostar molecules away from equilibrium using an internally driven microtubule-based active fluid. Activity lowers the critical temperature and narrows the coexistence concentrations, but only when there are mechanical bonds between the liquid droplets and the reconfiguring active fluid. Similar behaviors are observed in numerical simulations, suggesting that activity suppression of liquid-liquid phase separation is a generic feature of active LLPS. Our work describes a platform for building soft active materials with feedback control while also providing insight into cell biology, where phase separation emerged as a ubiquitous self-organizational principle.

**Introduction:** Liquid-liquid phase separation (LLPS) is a pervasive phenomenon that plays a central role throughout physics, materials science, engineering, biology, and everyday life[1–5]. In equilibrium, the concentrations of the coexisting phases are controlled by thermodynamic parameters such as temperature and pressure. The input of mechanical energy influences the properties of LLPS. Shearing a two-phase system can modify the coexistence concentrations and suppress the critical point[6–13]. Alternatively, systems can also be driven out of equilibrium not by the input of energy at macroscopic boundaries but by the continuous motion of energy-consuming microscopic constituents[14]. Such internally driven systems exhibit non-equilibrium dynamical states and phase transitions reminiscent of LLPS[15–18]. For example, even in the absence of attractive interactions, active Brownian particles undergo bulk phase separation into a dense liquid and a dilute gas [19–23]. In a different regime, increasing the motility of active particles with attractive interactions can suppress the phase separation and break up the dense condensate into clusters or a gel-like phase[24,25]. These observations identify the activity as a promising parameter that controls phase behaviors and can be easily patterned in space and time[26]. In active analogs of LLPS, such as Janus swimmers and motile bacteria, however, the same microscopic constituents both generate the active forces and undergo phase separation. Generalizing these findings to a much wider range of passive materials exhibiting LLPS remains a challenge.

We use activity to control LLPS of a passive system. Notably, activity can strongly couple to and deform diverse soft, passive materials, including liquid droplets, soft liquid interface, and lipid-bilayers[27–30]. Motivated by these observations, we study associative DNA nanostars that exhibit an equilibrium LLPS[31]. We merge this system with a microtubule (MT)-based active fluid exhibiting chaotic flows[32], finding that activity suppresses the critical point temperature, and narrows the concentration gap of the coexisting phases. The suppression of LLPS requires a direct mechanical link between the reconfiguring MT network and the passive entities undergoing the phase transition.

**Results**

**Coupling DNA-based droplets to an active fluid**: Our experimental system had three components. First, we used associative multi-armed DNA nanoparticles with attractive hybridization interactions at the tip of their arms. These four-armed molecules formed dense liquid-like condensates that coexisted with a dilute gas phase (**Fig. 1a, S1**)[31,33–35]. The second component was a network of filamentous MTs bundled by the depletion interaction[36]. MTs partitioned exclusively to the dilute gas phase of DNA nanostars. Finally, a fraction of the DNA nanostars was covalently linked with a kinesin molecular motor on one arm[37,38]. The kinesin-DNA motors had a dual purpose. When in the gas phase, they bound to MTs. They used energy from ATP hydrolysis to step along MTs, powering their interfilament sliding and generating chaotic active flows, similar to those previously studied[32]. The kinesin-DNA motors also dissolved in a dense liquid. When displayed on the surface they provided a bond between the continuously reconfiguring active network and the DNA-based droplets.

Passive DNA-based nanostars phase separated, forming spherical droplets that minimized surface tension. In comparison, activity significantly deformed the liquid droplets (**Fig. 1b**). In principle, such deformations could be driven by two interactions: short-ranged motor-mediated mechanical coupling of the reconfiguring MT network to the droplet or longer-ranged hydrodynamic flows generated in the dilute phase. To determine the dominant interaction, we removed the kinesin-DNA, thus severing droplet-filament bonds while keeping the same DNA composition. The active flows were powered by orthogonal streptavidin-biotin kinesin clusters. In this case, droplets were advected by the flow, but they largely retained their surface-tension-minimizing spherical shape (**Fig. 1d, Video S1**). This demonstrated that droplet-network coupling is the key interaction. Simultaneously visualizing droplets and MTs provided additional evidence for the importance of droplet-network coupling. In particular, the deformation of mechanically-coupled droplets followed the curvature of the MT bundles (**Fig. 1c, Video S1**). The active flows, combined with the connectivity of the droplet to the cytoskeleton network, changed the droplet's surface area. After reaching a critical extension, the droplet disintegrated into smaller droplets.

**Activity suppresses liquid-liquid phase separation**: The equilibrium phase diagram of associative DNA nanostars is temperature dependent[31,34]. At low temperatures, there was a phase separation between a dilute nanostar phase and dense spherical liquid-like droplets (**Fig. 2a**)[34]. With the temperature increasing above the critical point ($T_c = 24^oC$), the dense droplets evaporated, generating a uniformly dispersed fluid phase. We investigated how activity influenced the phase behavior for mechanically coupled droplets containing kinesin-DNA motors. At low temperatures $T < 19^oC$, we observed continuously deforming droplets coexisting with a dilute gas (**Fig. 2b, Video S2**). Above $19.5 \pm 0.5^oC$ the droplets evaporated. The difference in the critical-point temperatures between the two cases suggests that the non-equilibrium driving suppressed LLPS of DNA nanostars.

Next, we kept the temperature constant but controlled the concentration of kinesin-DNA, which increased the concentration of the microscopic force generators within the dilute phase and the mechanical coupling of the droplets to the network **(Fig. 2c).** At low kinesin concentration, the speed of the active flows was low, and droplets did not significantly deform **(Fig. 2c-e).** Increasing the kinesin-DNA concentration increased both the flow speeds, and the droplet deformation and reconfiguration. Above 5.5 µM kinesin-DNA concentration, the liquid-liquid coexistence disappeared, and we observed a uniform fluid phase.

Temporal dynamics provided further evidence for the interplay between activity and the LLPS. Active flows extract chemical energy from ATP hydrolysis. Once kinesin motors depleted the ATP fuel, the nonequilibrium dynamics ceased, and the active fluid speed dropped to zero (**Fig. 2g**). The decrease in the flow speeds was concomitant with the nucleation of new droplets and their subsequent growth (**Fig. 2f, S2, Video S3**).

**Measuring the influence of activity on the phase diagram**: The above-described experimental observations demonstrate that activity influences the LLPS of DNA nanostars. Motivated by these findings, we measured the phase diagram as a function of both temperature and activity. DNA concentration of the coexisting phases was estimated from the fluorescent intensities of the droplets and the background, and the droplet area fraction (**SI, Fig. S3**). Imaging liquid-liquid coexistence at different temperatures revealed the equilibrium phase diagram of DNA nanostars, with a binodal curve that separated the two-phase and the single-phase regime (**Fig. 3b**)[31]. Above a cloud temperature of $T_c = 23.5 \pm 0.3 ^oC$ there was only a uniform-density fluid phase.

To determine the influence of non-equilibrium driving, we measured the phase diagram in the presence of 5.5 $\mu M$ kinesin-DNA (**Fig. 3e, S1**). The number of DNA interacting arms influences the shape of the phase diagram[31]. To ensure meaningful comparison between phase diagrams we only changed one parameter. All samples had a total DNA-nanostar concentration of 65 $\mu M$, 75% had 4 overhangs, and 25% had 3 overhangs. To change the activity, we tuned the ratio 3 overhand nanostars with and without kinesin, while keeping their total concentration constant.

Activity generated by 5.5 $\mu M$ kinesin-DNA changed the phase diagram in two ways. First, it lowered the cloud point temperature to $18.3 \pm 0.3 ^OC$. Second, it decreased the concentration difference between the two coexisting phases. We repeated the same measurements for 3.0 $\mu M$ and 4.5 $\mu M$ kinesin-DNA. Combining these measurements uncovered a 3D non-equilibrium LLPS phase diagram that is a function of both temperature and activity (**Fig. 3a**). We also measured the phase diagram at $19 ^oC,$ and varied the kinesin-DNA concentration (**Fig. 3f**). Increasing kinesin concentration reduced the width of the tie-lines. The system did not phase separate beyond a critical kinesin-DNA concentration, which for $T = 19 ^oC$ was 4.5 $\mu M$. Importantly, the kinesin stepping speed is temperature dependent; the average speed of the active fluid increased ~120% when changing temperature from $19^oC$ to $23^oC$ [39].

Mechanically-coupled droplets were composed of 3 distinct elements: (i) DNA nanostars with four interacting arms, (ii) BG-modified DNA nanostars with three overhangs, (iii) kinesin motors attached to BG-modified DNA nanostars at a varying concentration $0 - 5.5\ \mu M$ (**Fig 1a**). Different DNA components could, in principle, partition differently between the two phases. For example, driven by their affinity for MTs, the kinesin-DNA could preferentially partition into the MT-rich dilute phase. At the same time, four-overhang nanostars could be enriched in the dense droplets. To check for this possibility, we labeled all three DNA-based components and determined their partitioning between the two phases. Importantly, all three components were distributed homogeneously within the droplet interior (**Fig. S4a-b**). Furthermore, all three components partitioned similarly between the dilute and dense phases (**Fig. S4c-d**). Thus, the total DNA concentration axis of the 3D phase diagram effectively described the partition of all three components **(Fig. 3a).**

In the dilute phase, nanostars could be monomeric or oligomerized due to their attractive ends. To characterize the possible structures, we used fluorescence correlation spectroscopy, a sensitive probe of the size of nanostar clusters. Such analysis revealed that the nanostars in the dilute phase were mainly either monomers or dimers, and there were no larger-sized clusters (**Fig. S5**). Interfilament sliding requires motor clustering. This activity might be generated by the DNA dimer population that can link multiple motors.

**Coupling between the active fluid and liquid droplets:** The activity-induced suppression of LLPS could originate either from the hydrodynamic interactions or the motor-mediated mechanical coupling between the droplets and the MT network. To discriminate between these two possibilities, we removed kinesin-DNA and generated activity using streptavidin-kinesin clusters that drove MT flows but eliminated mechanical coupling of MTs to droplets (**Fig. 4a**)[32]. Increasing the streptavidin-kinesin clusters increased active flows from 0 to $8\ \mu m/s$, up to 20-fold faster than the velocities generated at the highest kinesin-DNA concentrations. In the absence of mechanical coupling at the interface, the droplets remained circular at low velocities, while for higher velocities, they started deforming (**Fig. 4a**). However, increasing the velocity in this system did not alter the coexistence concentrations (**Fig. 4c**). Next, we kept the DNA concentration constant and changed the temperature. Again, there was no measurable difference between an equilibrium system and one driven by active flows but lacking MT-droplet coupling (**Fig. S6**). Adding 3 $\mu M$ of kinesin-DNA introduced filament-droplet coupling while minimally altering the speed of autonomous flows. In this case, coexistence concentrations narrowed, and critical point appeared (**Fig. 4b, c**). These measurements demonstrated the essential role of droplet-filament coupling for the activity-driven suppression of LLPS.

The non-equilibrium phase diagram demonstrates that essential role of motor-mediated droplets-MT coupling. To gain molecular insight, we simultaneously visualized the 3D MT network structure and the droplet shape. A projection reveals the structure of the MT network located up to 5.5 $\mu m$ from the droplet surface (**Fig. 5a, Video S4**). Droplets lacking kinesin-DNA components remained predominantly spherical, with MT bundles intermittently contacting their surface. The bundles did not generate coherent flows at the surface. In comparison, the presence of the kinesin-DNA link qualitatively changed the network-droplet coupling. Specifically, such droplets exhibited large shape distortions and spontaneous breakups (**Fig. 5b, Video S5**). Such deformations are driven by MT bundles that remain connected to the droplet surface over an extended time. Intriguingly, droplet breakup was driven by the MT alignment. The extensile motion of surface-bound MTs generated a narrow neck that eventually ruptured, creating two daughter droplets.

We quantified how the deformations generated in the dilute active fluid propagated into the interior of a dense droplet. Using particle tracking velocimetry of MT and bead-dopped samples, we measured the velocity field outside and within the droplets (**Fig. 5c, Video S6**). We extracted the eigenvalues of the simple shear rate, denoting the local deformation of each fluid element invariant to the coordinate system. The magnitude of the simple shear rate averaged over different contours as a function of the distance from the droplet surface revealed how deformations propagate across the interfacial boundarie (**Fig. 5d, Video S6**). Droplets lacking kinesin-DNA showed a profile quickly decaying to zero in the droplet interior, while kinesin-decorated droplets had a nonzero shear rate in the droplet.

**Continuum theory of LLPS in active fluid**: The activity control of the liquid-liquid phase diagram is captured by a minimal continuum model that incorporates the local concentration of the coexisting dilute gas and liquid DNA droplet phase $\phi$, the flow field of the active fluid $v_i$, and the nematic order parameter $Q_{ij}$. Similar models have been used before to describe the behavior of interfaces of active nematics and isotropic fluids[40–43], but the influence of activity on the phase diagram has not been studied. The temporal evolution of the fields is governed by the following equations

$$D_t \phi = \nabla^2 \left( \frac{\delta F_\phi}{\delta \phi} \right), \tag{1}$$

$$D_t Q_{ij} = \lambda D_{ij} + Q_{ik}\omega_{kj} - \omega_{ik}Q_{kj} + \frac{1}{\gamma}\frac{\delta F_Q}{\delta Q_{ij}}, \tag{2}$$

$$\rho D_t v_i = \mu \nabla^2 v_i - \partial_i P + f_i. \tag{3}$$

The concentration field, $\phi$, phase separates according to Cahn-Hilliard dynamics governed by the Landau-Ginzburg free energy $F_\phi$. $Q_{ij}$ relaxes at a rate controlled by the nematic free energy $F_Q$[14] and is stirred by the strain rate $D_{ij}$ and vorticity $\omega_{ik}$. The velocity field, $v_i$, is described by the Navier-Stokes equation with a forcing $f_i = f_i^\phi + f_i^a$ that consists of passive capillary forces $f_i^\phi$ and active forces $f_i^a = \partial_i \left( \frac{\phi}{\phi_+ + 1} \alpha Q_{ij} \right)$, with $\alpha$ an active stress, and $\phi_+$ the dilute phase concentration. The form of the free energy and the active forcing ensures that activity is confined to the dilute phase of $\phi$ (SI).

Extensile active stresses ($\alpha < 0$) generate spontaneous flows in the dilute phase [14,44,45], breaking up a bulk phase-separated system into finite-size droplets. This is analog to the effect of an externally applied shear that is known to mix phase separated mixtures[8,46], although in our system the stirring is driven by local shear flows generated spontaneously by active processes in the MT liquid crystal and transmitted to the droplets interior. At low activities, the coexisting phases have a finite density difference (**Fig. 6b, Video S7**), which decreases with increasing activity and eventually reduces to zero (**Fig. 6a, Video S7**). To quantify these observations, we map the phase diagram as a function of activity $\alpha$, reduced temperature $a \propto \frac{T-T_c}{T_c}$, and the volume fraction $\phi_0$. We observed that activity shifts the critical point $a_c$ to lower temperatures and to lower values of $\phi_0$, rendering the coexistence curve asymmetric about $\phi_0 = 0$. The resulting phase diagram reproduces many of the features observed in experiments (**Fig. 6c**).

Our model does not contain an explicit mechanical linkage between active fluid and liquid droplets. However, such coupling is ensured by using a single velocity field that varies continuously across the fluid-fluid interface. As seen in the experiments, shear flows generated in the active fluid phase propagate into the droplet interior, demonstrating that the single velocity field provides a proxy for mechanical coupling at the interface (**Fig. S7**). The current model does not allow one to turn off the coupling while maintaining the bulk activity. Therefore, we cannot verify that the mechanical coupling of DNA and MTs is essential for the suppression of phase separation. This would require a two-fluid model of an active MT liquid crystal and a DNA fluid, with different flow velocities and materials parameters and possible flow slip at the interface. Additionally, the viscoelasticity of the DNA droplets may play a role in suppressing the penetration of active flows in the absence of mechanical coupling and, therefore likely needs to be included.

**Conclusions**: We merged an active fluid with an equilibrium LLPS to attain activity control of the resulting non-equilibrium phase diagram. The stresses induced by the reconfiguring MTs translocate, rupture, and dynamically disassemble DNA-based liquid droplets, changing the liquid-liquid coexistence concentrations and suppressing the critical point temperature. From a fundamental perspective, these results raise intriguing questions about similarities and differences in how internally generated and externally imposed flows affect the equilibrium phase separation. From a material science perspective, active LLPS provides a foundation for assembling diverse soft active materials. In particular, the coupling between phase separation and activity open the path towards creating materials that can incorporate feedback and maintain homeostasis. Finally, our results are also relevant to cell biology, where phase separation has emerged as an important self-organizing principle[4,5,47]. Biological phase separation processes are usually described using thermodynamic formalisms that assume local equilibrium[48]. The cell cytoskeleton, however, is collectively driven away from equilibrium by nano-scale force-generating processes. Our results suggest that active cytoskeletal forces could locally control the spatiotemporal dynamic of membraneless organelles, but this likely requires mechanical coupling between the cytoskeletal network and the liquid droplets[49,50].

**Acknowledgments:** We thank Paarth Gulati, Zhihong You, Gabriel Abrahams, and Sho Takatori for fruitful discussions. We thank Noah Mitchell for help with the analysis of 3D images. The experimental work was supported by Department of Energy, Basic Energy Sciences through award

DE-SC001973. The development of the theoretical model was supported by NSF-DMR-2041459. We also acknowledge the use of the Brandeis MRSEC Bio-Synthesis Facility, which is supported by NSF-MRSEC-2011846. O.S. also acknowledges support from the W.M. Keck Foundation. A.M.T. is a Simons Foundation Fellow of the Life Sciences Research Foundation and is an Awardee of the Weizmann Institute of Science–National Postdoctoral Award Program for Advancing Women in Science.

**Figures**

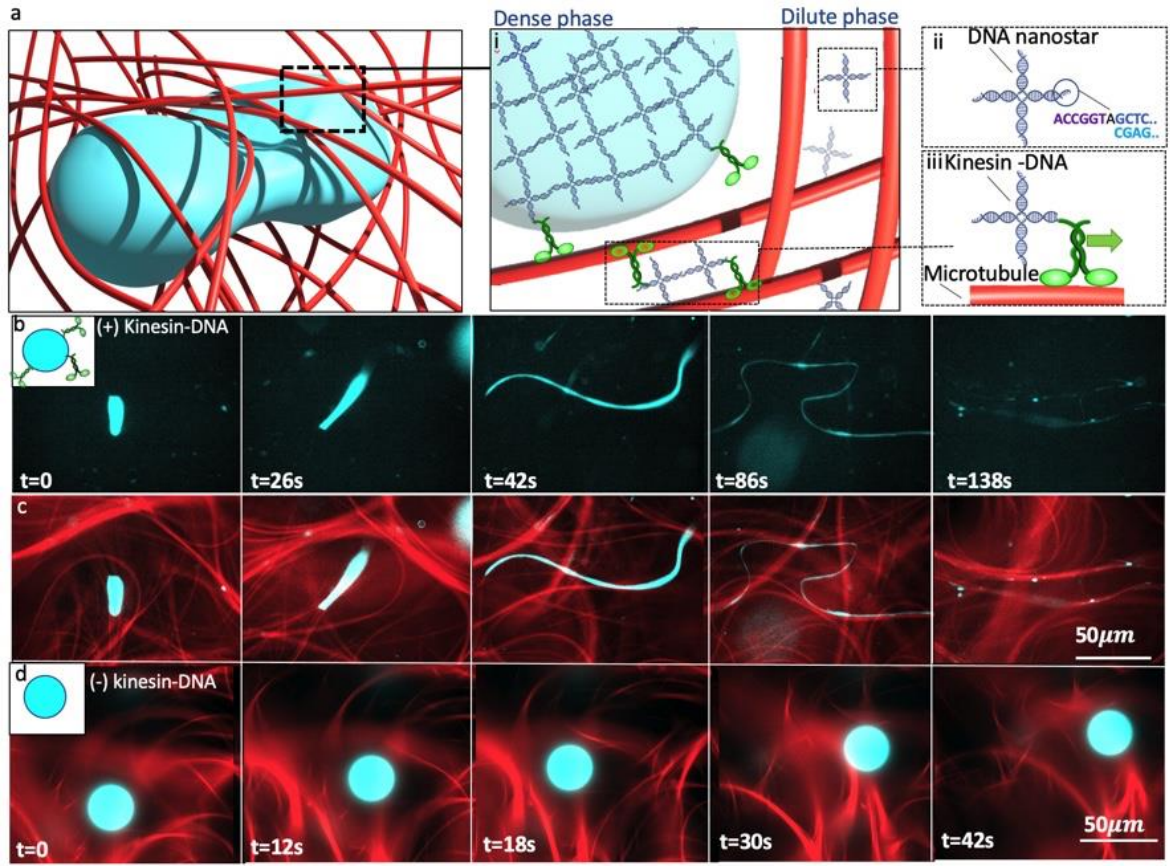

**Fig. 1 – Active-LLPS composed of DNA nanostars and an MT-based active fluid. a)** DNA droplet enveloped by an active MT network. DNA nanostars coexist between dense droplets and dilute gas, driven by the hybridization of a palindromic sequence at their tips. MTs partition into the dilute phase. Kinesin-DNA has a kinesin motor attached to one arm, while the other three arms have attractive interactions. They bond droplets to MTs, and in the dilute phase generate active flows. **b)** DNA droplet advected and deformed by the active MT network. **c)** Overlaying DNA droplets (cyan) and MTs (red) demonstrates coupling between the droplets and the MT network.

**d)** Replacing kinesin-DNA with streptavidin-based clusters eliminates droplet-network coupling, while preserving activity. Droplets are not significantly deformed but are advected by the active flows.

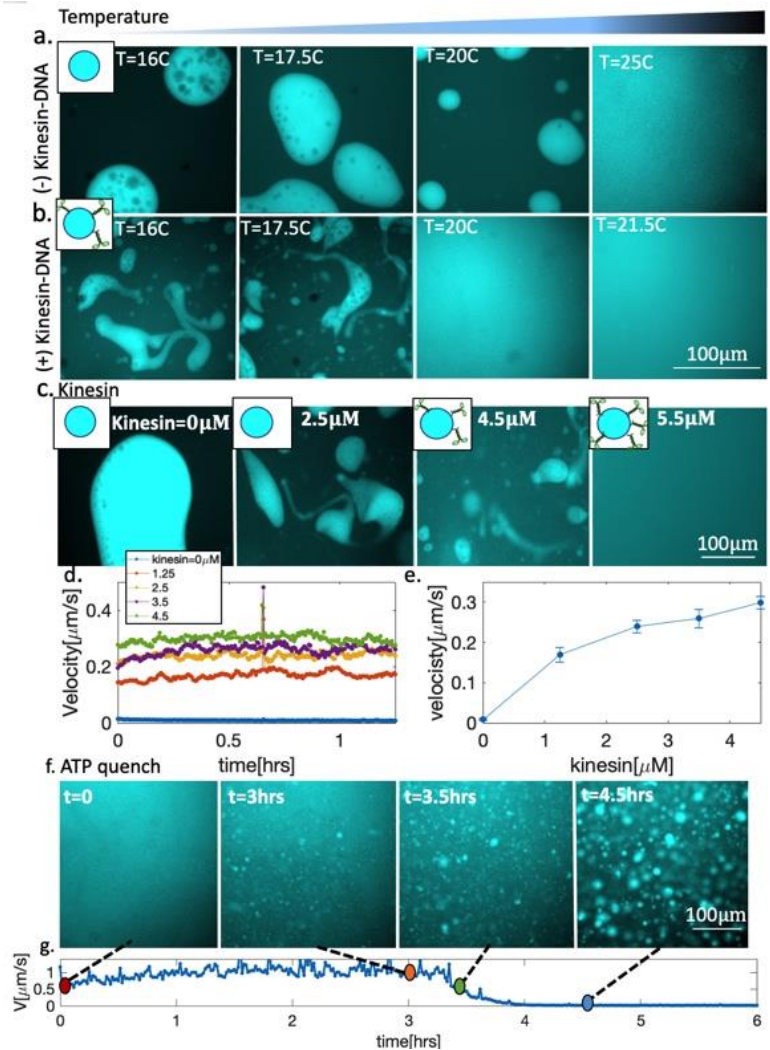

**Fig. 2 – Activity suppresses LLPS. a)** DNA droplets coexisting with an active fluid at different temperatures. The sample lacks kinesin-DNA, and activity is generated by kinesin-streptavidin motor clusters. **b)** DNA droplets coexisting with an active fluid in the presence of kinesin-DNA, which links MTs to droplets. **c)** Droplets with increasing kinesin-DNA concentration which increase both activity and network-droplet coupling. **d)** Steady-state flow speeds flows generated at different kinesin-DNA concentrations. T=$18.5^oC$. **e)** Active flow speeds as a function of kinesin-DNA concentration. **f)** DNA droplets during an activity quench. Depleting ATP reduces

flow speeds, resulting in droplet nucleation, growth and coarsening. **g)** Speed of active flows over the sample lifetime.

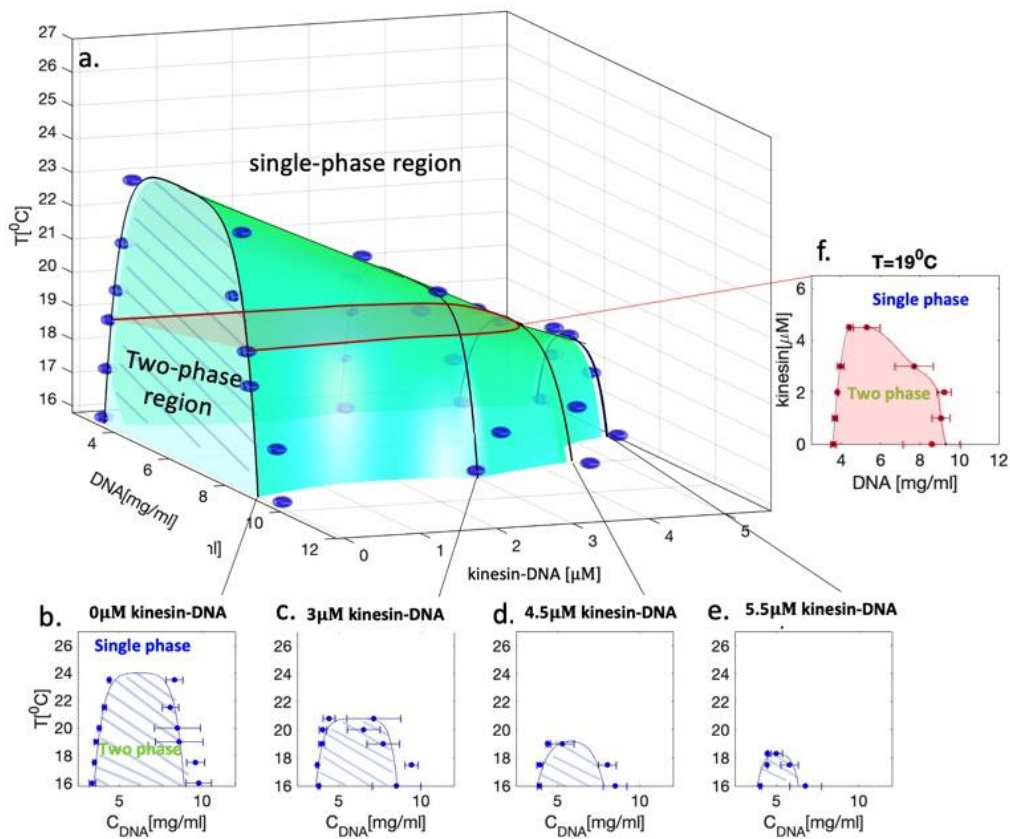

**Fig. 3 - Activity-dependent phase diagram of DNA nanostars.** (**a**) LLPS phase diagram as a function of both temperature and activity (**b-e**) Activity is controlled by tuning kinesin-DNA concentration between 0-5.5 $\mu M$. (**f**) Phase diagram at a function of kinesin-DNA concentration at constant temperature. Error bars are standard deviation, $n = 3 \vee 4$. Black lines are eye guides. The surface is an interpolation between the black lines.

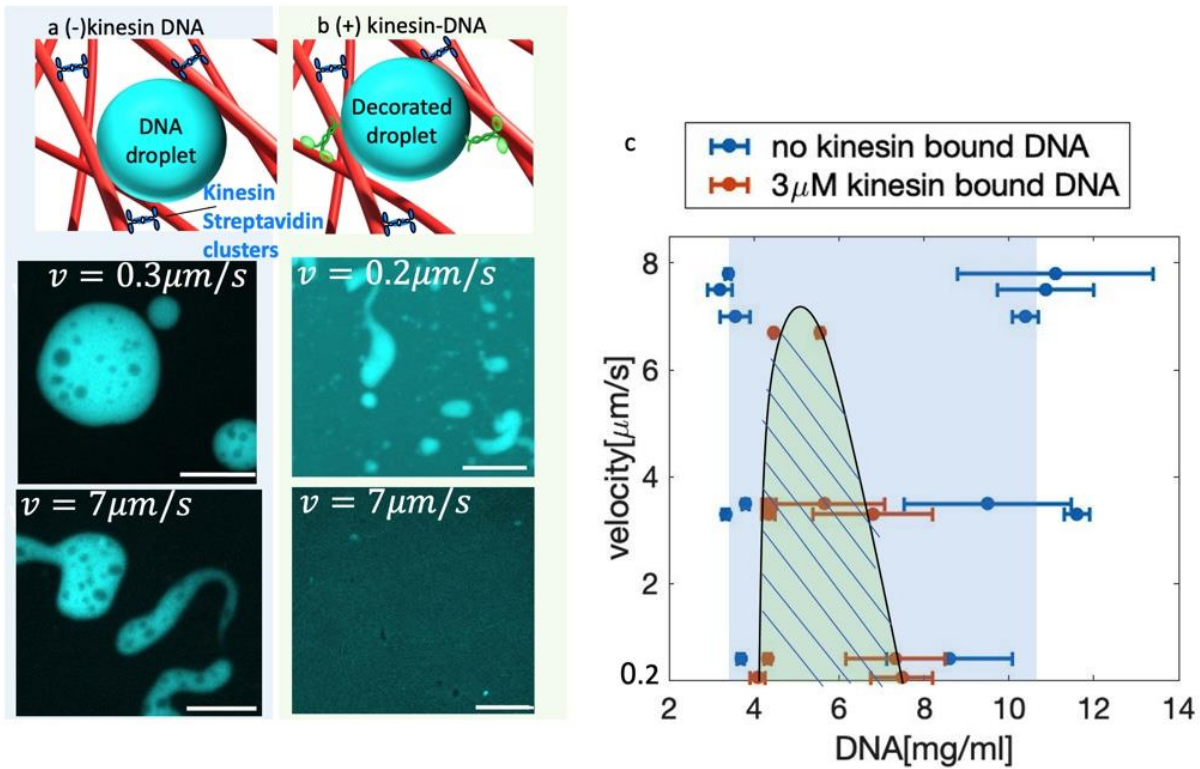

**Fig. 4 – Suppression of LLPS requires droplet-MT interactions. a)** Droplets lacking kinesin-DNA where active is controlled by streptavidin-kinesin clusters. T=$19^oC$. **b)** DNA droplets were observed at two flow velocities controlled by increasing the concentration of streptavidin-kinesin clusters. The sample had 3 μM concentration of kinesin-DNA, ensuring the droplet-filament coupling. **c)** Phase diagram as a function of the velocity for a system lacking kinesin-DNA (blue) and in the presence of 3 μM kinesin-DNA. Velocity is controlled by streptavidin-kinesin clusters. Error bars are the standard deviation, $n = 3 \vee 4$. Lines are a guide to the eye.

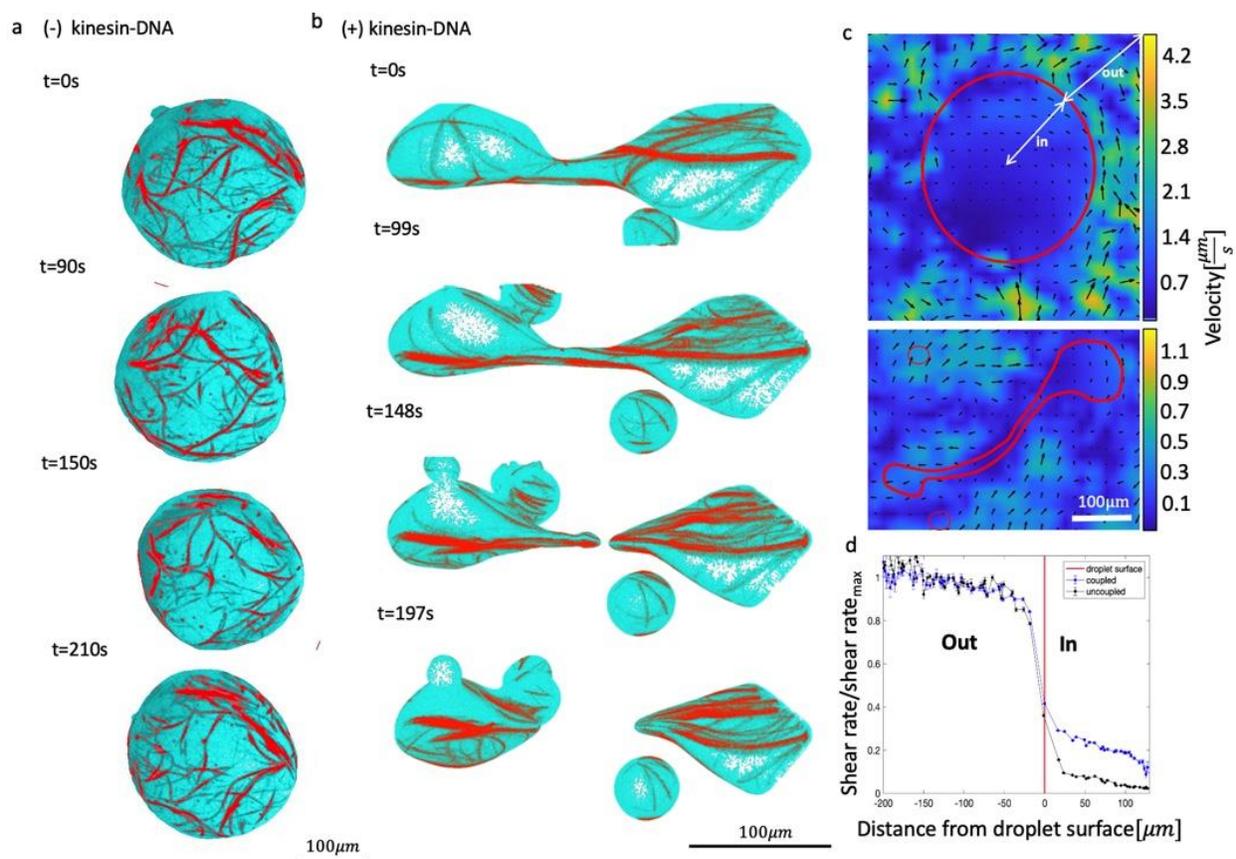

**Fig. 5 – Coupling of MT network to DNA droplets. a)** MT-droplet interactions in a system lacking kinesin-DNA. Droplet is in cyan only MTs within 11 $\mu m$ of the droplet surface are shown. **b)** Coupling between DNA droplets and MT network in the presence of kinesin-DNA. Kinesin-DNA concentration is 3 $\mu M$. MTs within 5.5 $\mu m$ of the droplet surface are shown **c)** Particle image velocimetry of 200 nm particles embedded in DNA droplet for (left) passive droplet (right) DNA decorated droplet. **d)** Normalized simple shear rate dependence of the distance from the droplet edge. Each point is binned over 400-time points and averaged over 3 samples. Error bars are the size of the data points.

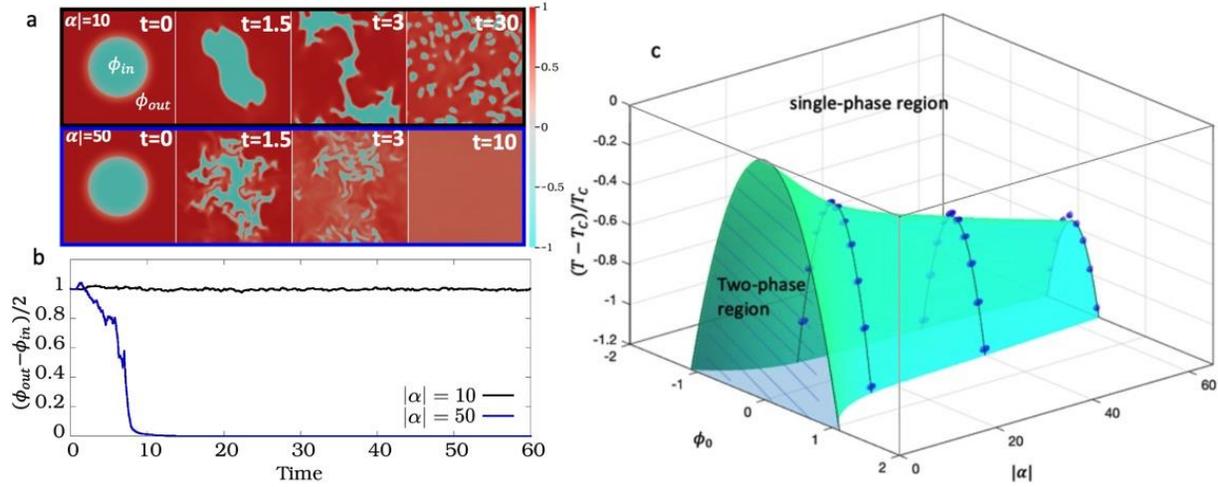

**Fig. 6 - Theoretical model of active LLPS. a)** Time-lapse images obtained from simulations for two different activity values $|\alpha| = 10$ (top) where activity arrests, but does not suppress phase separation, and $|\alpha| = 50$ (bottom) where activity suppresses phase separation mixing the system into a homogeneous fluid. Both are below the critical reduced temperature $a = -1$ of the passive system. **b)** The difference between the coexisting concentrations as a function of time for the two runs. A value of $\delta\phi = 1$ corresponds to phase separation while $\delta\phi = 0$ corresponds to a uniform state. **c)** Activity-dependent liquid-liquid phase diagram. The curve for $|\alpha| = 0$ is the mean-field coexistence region of the Landau-Ginzburg free energy. The surface is an interpolation between the black curves.

## Supplementary information for:
## Controlling liquid-liquid phase behavior with an active fluid

**Methods**

**Tubulin purification and MT polymerization:** Tubulin was purified from bovine brain through 2 cycles of polymerization and depolymerization[1]. Tubulin was stored at $-80^oC$ and recycled through an additional cycle of polymerization and depolymerization. Tubulin was labeled with 647-Alexa dye for fluorescent imaging using a succinimidyl ester linker[2]. Recycled and labeled tubulin was flash-frozen in liquid nitrogen and stored at $-80^oC$. MTs were polymerized at a tubulin concentration of 8 mg/ml in an M2B buffer (80 mM Pipes pH=7.0, 1 mM EGTA, 2 mM MgCl2) containing 20 mM DTT and 10 mM GMPCPP (Jena bioscience), a non-hydrolysable analog of GTP. The mixture was incubated on ice for 10 minutes, subsequently at $37^oC$ for 30 minutes, and allowed to sit at room temp for 3 hours, flash-frozen in liquid nitrogen, and stored at $-80^oC$. Fluorescent microtubules contained 5% Alex-647 labelled tubulin.

**Kinesin purification:** Dimeric kinesin, consisting 401 amino acids of the N-terminal motor domain of *D. melanogaster* kinesin-1 fused to the SNAP-tag, was purified as previously described[3,4]. The SNAP-tag is appended to the cargo binding region of the motor. The protein was flash-frozen in liquid nitrogen and stored at $-80^oC$.

**Modifying DNA oligos:** 5′-amine-modified DNA oligos (Integrated DNA Technologies) were labeled with BG-GLA-NHS (New England Biolabs)[5–7]. Briefly, BG-GLA-NHS was dissolved in DMSO to a final concentration of 20 mM. Amine-modified DNA oligos were dissolved at 2 mM in HEPES buffer (150 mM, pH=8.4). BG-GLA-NHS solution was added to DNA oligos at a final BG concentration of $7~\mu M$. The mixture was incubated for 30 min at room temperature. DNA was separated from excess BG using a size exclusion spin column (Micro Bio-Spin 6 columns, Bio-Rad). Before DNA cleaning, tris buffer in the column was exchanged with phosphate-buffered saline (0.2x PBS) (pH=7.2) according to manufacturer instructions. The separation step was repeated four times. DNA oligos were concentrated seven times the column volume using a speed vac to a final concentration of $5 - 10~\mu g/\mu l$.

**Assembling DNA nanostars:** Four-armed DNA nanostars were assembled from four distinct 49-bp long DNA oligos[8], with 6-bp sticky palindromic overhang AGGCCT, and one unpaired base (**Fig. S1**). DNA oligos were mixed at an equal concentration of $20~\mu g/\mu l$ in TAE buffer (40 mM Tris, 20 mM acetic acid, 1 mM EDTA, pH=8.2), 25% of sequence number 1 were labeled with a BG modification. DNA oligos were annealed to their complementary strands. DNA was heated to $95^oC$ for 10 min and gradually cooled down to room temperature in a heat block left on the bench. DNA nanostars were stored at −20 °C.

**DNA nanostar oligos sequence:**

Sequence-1: AGGCCTAGCTCAACGGTGAGTTGCACGTTCGGCCGTTCCCAGAGAAAGC
Sequence-1: for modifications
AGCTCAACGGTGAGTTGCACGTTCGGCCGTTCCCAGAGAAAGC
Sequence- 2: AGGCCTAGCGTCAAACCACACTCGCACTTCGTGCAACTCACCGTTGAGC
Sequence- 3 -AGGCCTAGCCGCCGCCGCATCGCCCGCTTGTGCGAGTGTGGTTTGACGC
Sequence- 4 -AGGCCTTGCTTTCTCTGGGAACGGCCGTTGCGGGCGATGCGGCGGCGGC

**ATP-regeneration and anti-oxidant mixture:** The samples contained an ATP-regeneration and antioxidant mixture which ensured steady-state kinesin stepping speed while minimizing photobleaching effects. The final sample contained 26.6 mM phosphoenolpyruvic acid (PEP), 1.4 mM ATP, 6.7 mg/mL glucose, 0.4 mg/mL glucose catalase, 0.08 glucose oxidase, 5 mM MgCl2[9].

**Binding DNA nanostars to kinesin motors:** Modified DNA nanostars were thawed at $45^oC$ for 5 minutes and cooled down to room temp. The mixture was added at equal volume to the ATP regeneration and antioxidant mixture and incubated for 2 hrs at $15^oC$. Kinesin motors stock was added to the mix at the desired concentration and mixed using a pipette with a wide tip. The mixture was incubated at $15^oC$ for 1 hr. Previous work estimated the efficiency of kinesin- DNA binding at 70-80%[7].

**Assembling active gel:** Polymerized MTs were added to the ATP-regeneration system at a final concentration of 2 mg/ml. The MT mixture was added to the modified kinesin decorated DNA nanostars at a ratio of 2:5. The mixture was introduced in an acrylamide-coated flow chamber with dimensions of 0.12× 2.0 × 20.0 mm, sealed with Norland Adhesive glue, and cured with UV[2].

**Fluorescent labeling of kinesin:** SNAP tagged kinesin motors were mixed with BG-Alexa488 (New England Biolabs), at a ratio of 10:1, in M2B buffer containing 0.5 μM DTT. The mixture was incubated for 30 minutes at room temp and sequentially added to the BG labeled DNA nanostars.

**Optical microscopy imaging:** For imaging and quantifying total DNA concentration we used YOYO-1 intercalating dye at a concentration of 100 to 200 nM. For measuring-tie lines, DNA nanostars were internally labeled with Cy5 fluorophore at the core of the nanostar. Active gels were imaged using conventional fluorescence microscopy Nikon Ti-2 and an Andor-Zyla 5.5 camera running open-source microscopy managing software Micro-Manager 1.4.23. Three-dimensional volumes and droplet deformation profiles were extracted from confocal imaged acquired with a Nikon eclipse Ti-2 with CrestOptics Confocal X-light V2 and a Photometrics 95b camera. The experiments were conducted on a Tokai HIT CBU thermal plate operating at a range of $10 - 55^oC$. FCS measurements were conducted on a Leica SP8 confocal scanning laser microscope, with a Plan Apo 63x/1.2 N.A. water immersion lens and the Leica FCS module.

**Measuring DNA concentrations between coexisting phases:** Partitioning of DNA between the dilute and dense phases was calculated from the fluorescent intensity and the 2D area fraction of each phase. Since the droplets diameter is typically larger than the chamber height of 120 μm, the active-LLPS were treated as quasi-2D sample, therefore the ratio of the area fraction indicates the ratio of the volume fractions. To measure the intensity of the entire volume section we used a 2x/0.1 NA objective with a 100 μM depth of field. Close to the binodal curve the droplets diameters were smaller than the channel depth, therefore area volume fraction of smaller droplets was overestimated. Each measurement was kept at a constant temperature until the ATP was depleted. For the active-LLPS the concentrations were extracted from steady state values of each phase concentrations about an hour before the ATP depletion (**Fig. S5**). For the passive regimes, the measurements were taken from after droplets relaxed to their final shape at about t=10-15 hr.

**DNA cluster size in the dilute phase:** We characterized the structure of the dilute phase using fluorescence correlation spectroscopy. The fluorescent intensity fluctuations due to the Brownian motion reveal the distribution of diffusion times and consequently the size of nanostar clusters. Fluorescent correlation spectroscopy measurements were conducted in the dilute DNA phase for kinesin decorated DNA droplets and for passive DNA droplets.

**Projection of MTs on DNA droplets:** To quantify the interactions between the DNA and the MTs, 3d confocal images were acquired. The evolving surface of the droplets was extracted from the 3d volume images using TUBULAR toolkit, with a MT projection at a distance of $5 - 11\ \mu m$ from the surface[10].

**Theoretical model for active liquid phase separation:** The binary mixture is described by three continuum fields: a phase field $\boldsymbol{\phi}$, a nematic field $\boldsymbol{Q_{ij}}$, and a velocity field $\boldsymbol{v_i}$. The equations of motion are given by:

$$D_t \phi = M \nabla^2 \left( \frac{\delta F_\phi}{\delta \phi} \right) \tag{1}$$

$$D_t Q_{ij} = \lambda D_{ij} + Q_{ik}\omega_{kj} - \omega_{ik}Q_{kj} + \frac{1}{\gamma}\frac{\delta F_Q}{\delta Q_{ij}} \tag{2}$$

$$\rho D_t v_i = \eta \nabla^2 v_i - \partial_i P + f_i \tag{3}$$

The phase field $\phi$ represents the local concentration of the two species that phase separate in equilibrium, advected by the flow $v_i$, with $D_t = \partial_t + v_i \partial_i$. The field $\phi$ evolves with diffusive dynamics of the Cahn-Hilliard type, with $F_\phi$ the phase field free energy, given by

$$F_\phi = \frac{1}{2}\int d^d r \left[ a\phi^2 + \frac{b}{2}\phi^4 + k(\nabla \phi)^2 \right]. \tag{4}$$

The parameter $a \sim T - T_c$ controls the transition, with $a = 0$ the critical point. The demixed state corresponds to $a < 0$. In equilibrium, this describes phase separation in two fluids of bulk densities $\phi_\pm = \pm\sqrt{-a/b}$. Since $\phi$ represents the local density, we can interpret one of the bulk values, for instance $\phi_+$, as the value corresponding to full concentration of active MT liquid crystal, and $\phi_-$ as the value corresponding to full concentration of the passive fluid (here the DNA condensates). Since mass is conserved, the total density $\phi_0 = \int d^d r \phi(r,t)/L^2$ is constant, where $L$ is the system size.

The nematic field $Q_{ij}$ captures local orientational order of the MT bundles. Its dynamics is governed by active nematic hydrodynamics[11] that couples flow gradients described by the symmetrized strain rate $D_{ij} = (\partial_i v_j + \partial_j v_i)/2$ and vorticity $\omega_{ij} = (\partial_i v_j = \partial_j v_i)/2$ to relaxational dynamics described by a Landau-de Gennes free energy

$$F_Q = \frac{1}{2}\int d^d r \left[\frac{a_Q}{2}(\tilde{\phi}+1) Tr\mathbf{Q}^2 + \frac{b}{4}[Tr\mathbf{Q}^2]^2 + \frac{k_Q}{2}(\partial_i Q_{jk})^2\right]. \quad (5)$$

The parameter $a_Q$ is weighted by $(\tilde{\phi}+1)/2$, with $\tilde{\phi} = \frac{\phi}{\phi_+}$, so that nematic order is built only in the regions of the system where the field $\phi$ describes the presence of an active nematic.

Finally, the flow $v_i$ follows a Navier-Stokes equation in which the forcing term $f_i$ has two components, $f_i = f_i^\phi + f_i^Q$. The first is an equilibrium term produced by density gradients, written in terms of the chemical potential $f_i^\phi = \phi \partial_i \left(\frac{\delta F_\phi}{\delta \phi}\right)$[12]. This term can be rewritten as the divergence of a stress[13], $f_i^\phi = \partial_j \sigma_{ij}$, with $\sigma_{ij} = -k\left(\partial_i \phi \partial_j \phi - \frac{\delta_{ij}(\nabla \phi)^2}{2}\right)$ in two dimensions. The second forcing term is the gradient of the active stress that couples to liquid crystalline degrees of freedom, $f_i^a = \partial_j \sigma_{ij}^a$, with $\sigma_{ij}^a = \frac{\alpha}{2}(\tilde{\phi}+1)Q_{ij}$. The parameter is determined by the biomolecular processes that drive active dipolar forces, which are taken to be extensile ($\alpha < 0$). The factor $\tilde{\phi}$ guarantees that activity can only be present in the nematic.

Related models were used to study interfaces between active nematics and isotropic fluids, as well as chemotaxis [14–17]. These systems have, however, been studied in the limit of low activity limit where the dynamics of $Q_{ij}$ is slow, and the phase diagram has not been previously analyzed. When $\alpha = 0$ the model captures equilibrium phase separation via Cahn-Hilliard dynamics due to the passive immiscibility of the two species. Increasing extensile activity, generates active spatio-temporal chaotic flows in the bulk[11,18,19] that can render the equilibrium interface unstable. These turbulent flows arising from coupling to liquid crystalline degrees of freedom in the bulk of the active nematic provide sustained local stirring and fight against the diffusive dynamics, mixing the fluid mixture into a uniform state.

Bulk active stresses that couple to liquid crystalline degrees of freedom are essential to yield a shift of the critical temperature. We have verified that interfacial active stresses that couple to gradients of $\phi$, as used in active models B and H[20,21], do not suppress the critical point. In other words, in a purely scalar model time reversal symmetry-breaking terms can arrest the phase separation, but cannot suppress it[22–24]. This is because these interfacial stresses act only at the interface and are not able to stir continuously the bulk of the fluid.

We have shown numerically that bulk active flows provide a self-stirring force capable of fully mixing the system. This was quantified by examining the behavior of $\delta\phi = \frac{\phi_{max}-\phi_{min}}{2}$ as a function of activity $\alpha$ and reduced temperature $a$, where $\phi_{max}$ and $\phi_{min}$ are the bulk values of the concentration in each phase at a given time. The quantity $\delta\phi$ represents the width of the coexistence region, therefore the critical point can be approximately identified with the reduced temperature for which $\delta\phi = 0$.

The numerics have been carried out with the software package OpenFOAM to integrate the Navier Stokes equation using the PIMPLE algorithm, while integrating the equations for $\phi$ and $Q_{ij}$ with Euler steps[25]. We have scanned different values of $a$ and $\phi_0$ for three values of activity. The points for which $\delta\phi$ vanishes are the points of the coexistence curve that create the phase diagram of Fig. 6.

Experimental activity is controlled by the DNA-kinesin concentration, which also increases the coupling between the MTs and DNA droplets Equations (1-3), however, do not contain an explicit mechanical coupling, as this would involve several interacting flows. We have simplified this description by having a single flow $v_i$, and by tuning activity through the strength of the active stress $\alpha$. Having a single flow means the two phases of $\phi$ are effectively coupled for any value of activity; indeed, if a volume $V$ contains a mixture of both species, represented by $\phi_- < \phi < \phi_+$, a single flow will advect both species in the same way (since it just advects $\phi$). Having a single flow in the model that serves as an effective coupling between the two fluids means that it is not possible to switch off the mechanical coupling, while maintaining finite activity in the bulk, and verify that this eliminates the suppression of phase separation, as has been done in the experiments.

**Theoretical Phase diagram parameters:** The data points are obtained from a simulation by scanning values of reduced temperature $a$ and inspecting $\delta\phi$, and the black lines are parabolic fits for each value of $\alpha$T. he simulation is run on a lattice of size L = 10, divided in $128 \times 128$ cells, with $\eta = 1, \gamma = 1, a_Q = -b_Q = -1, \lambda = 1, M = 0.1, a = -b = -1$ and $k = 0.004$. Units of time are set by $\gamma$, units of length by $L$, and $a_Q = b_Q = 1$ which sets the units of stress.

**References**
1. Castoldi, M. & Popov, A. V. Purification of brain tubulin through two cycles of

23. Cates, M. E. & Tailleur, J. Motility-Induced Phase Separation. *Annu. Rev. Condens. Matter Phys.* **6**, 219–244 (2015).
24. Cates, M. E., Marenduzzo, D., Pagonabarraga, I. & Tailleur, J. Arrested phase separation in reproducing bacteria creates a generic route to pattern formation. *Proc. Natl. Acad. Sci.* **107**, 11715–11720 (2010).
25. Moukalled, F., Mangani, L. & Darwish, M. *The Finite Volume Method in Computational Fluid Dynamics*. vol. 113 (Springer International Publishing, 2016).
26. Han, C. C., Yao, Y., Zhang, R. & Hobbie, E. K. Effect of shear flow on multi-component polymer mixtures. *Polymer (Guildf).* **47**, 3271–3286 (2006).
27. Zadeh, J. N. *et al.* NUPACK: Analysis and design of nucleic acid systems. *J. Comput. Chem.* **32**, 170–173 (2011).
25

**Supplementary Figures:**

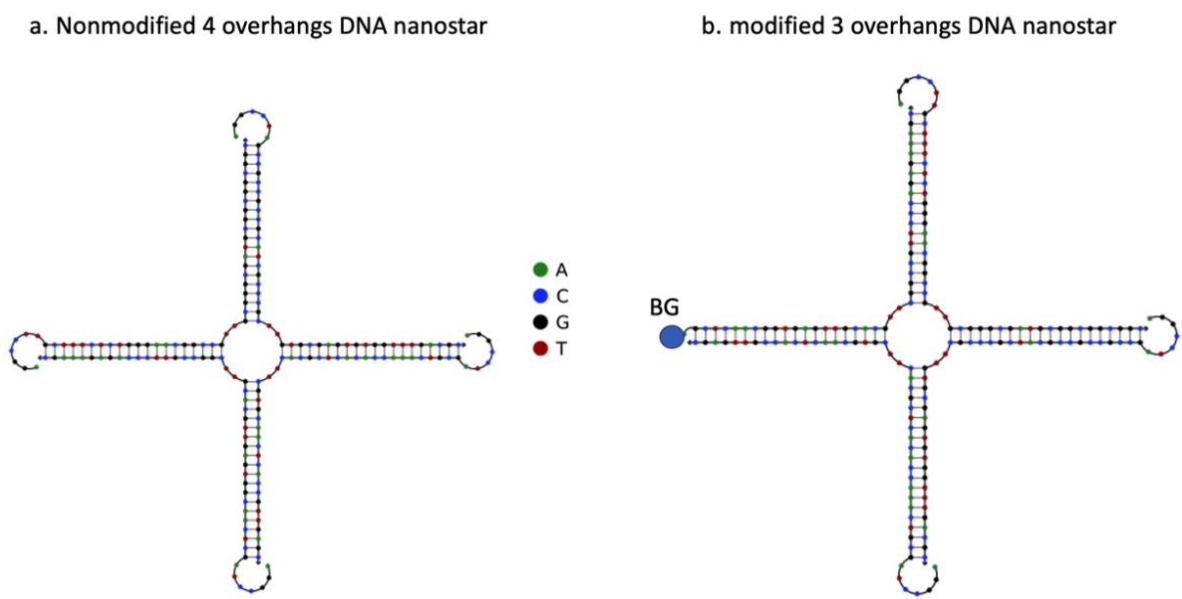

**Figure S1: Structure of DNA nanostars**. (a) Four-armed DNA nanostar with single-stranded overhangs at each tip. Attractive interactions between tips drive LLPS. (b) DNA nanostar modified with a benzo-guanine (BG), and no overhang on the modified arm. The BG tag forms a covalent bond with SNAP tag that is fused to the kinesin motor. The structures are visualized with NUpack[27]

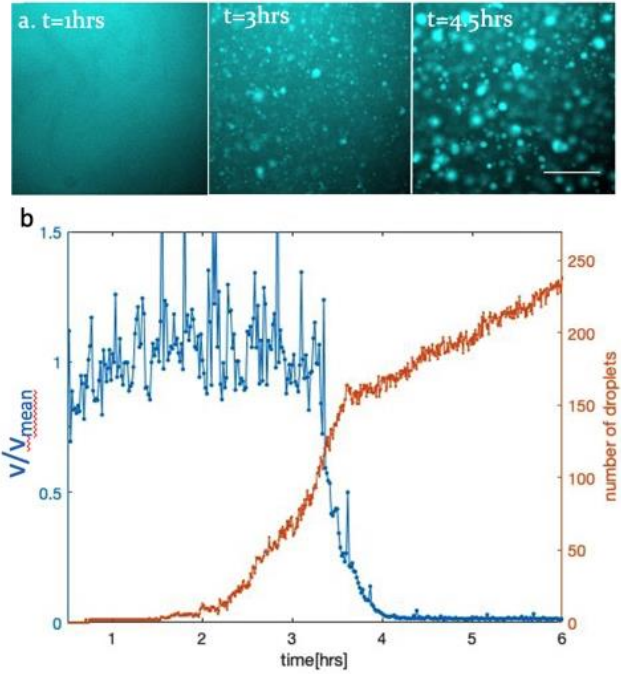

**Figure S2: ATP depletion leads to droplet nucleation**. (a) Image of additional DNA droplet forming as the active gel slows down, running out of ATP, at $T = 18.5^oC$, and kinesin concentration of $k = 3\ \mu M$. Scale bar $100 \mu m$. **(b)** MT gel velocity as a function of time (blue), and number of DNA droplet in the field of view as a function of time (red).

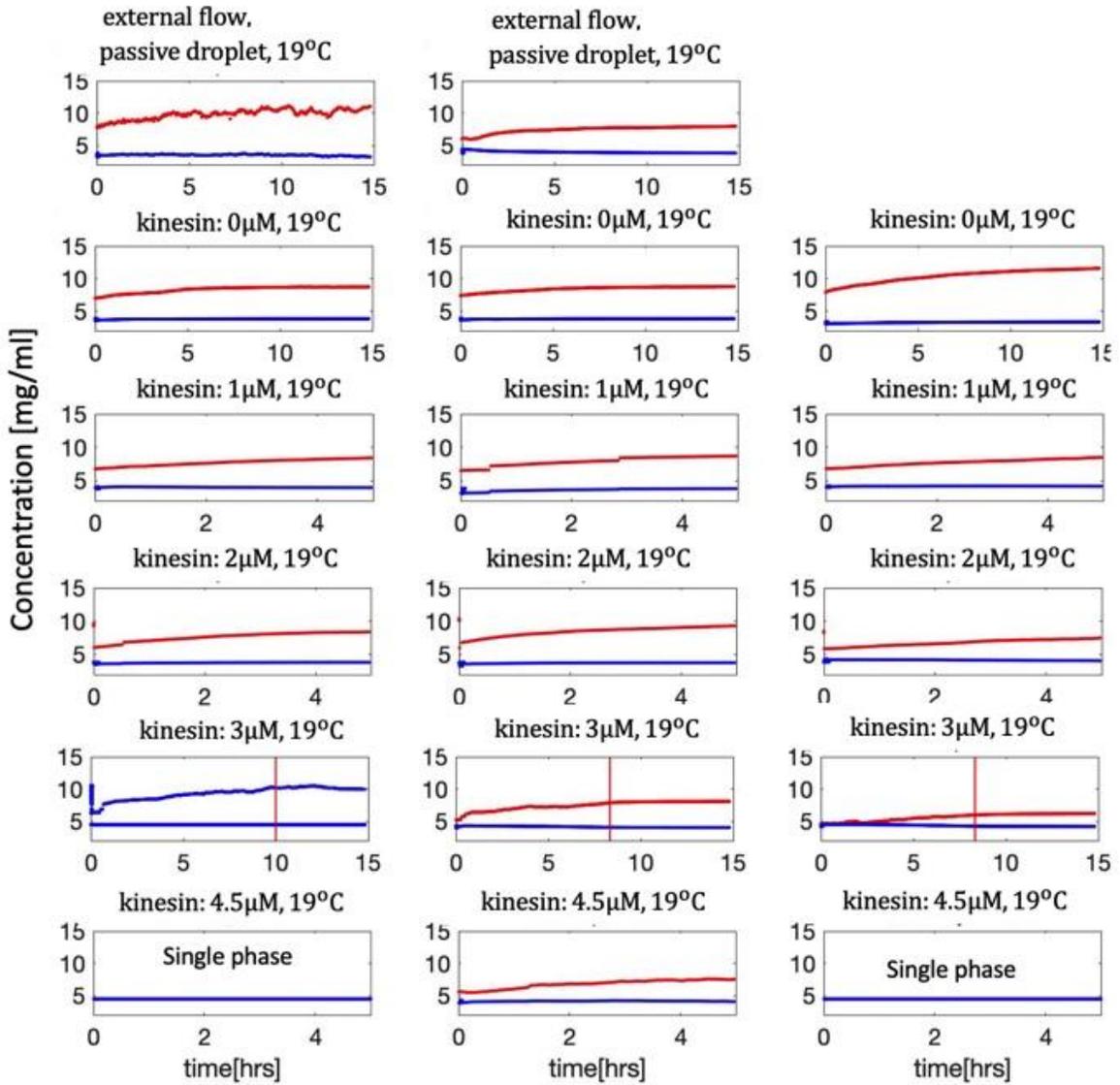

**Figure S3: Temporal dynamics of active-LLPS.** Concentration of dilute (blue) and dense (red) phase calculated from fluorescent values as a function of time for different sample conditions. Red orthogonal line indicates sample velocity reaching $v = 0$. Single blue curve indicates a single phase. Values of concentrations figure 3 are taken as steady state values 1-1.5 hrs before sample death. Temperature is $19^oC$.

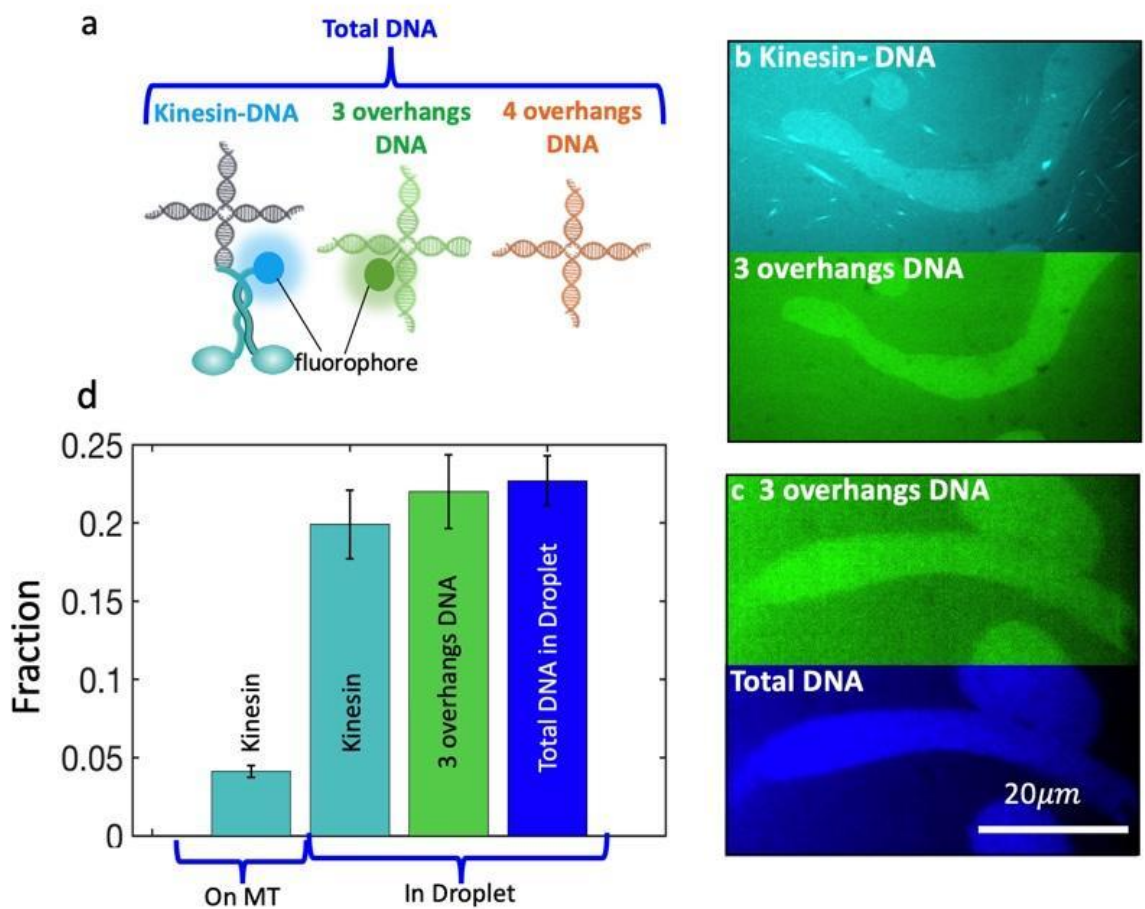

**Figure S4: Partitioning of different DNA species across coexisting phases. a)** Schematic of three different DNA nanostars present in active-LLPS. **b)** Images of fluorescently labeled kinesin-DNA and of 3 overhang-DNA nanostar shows their uniform spatial distribution with the dense droplets. Ten percent of kinesin motors were labeled with a SNAP-Alexa 488 prior to binding to the DNA nanostar. 3-overhangs DNA nanostars were labeled with Atto-647 fluorophore at its core. **c)** Image of 3-overhangs nanostar and total DNA distribution within . DNA was labeled with YOYO dye. **d)** Fraction of the different nanostar species found in the droplets. Kinesin-DNA is present both in the dilute phase and bound to microtubule network. Error is standard error averaged over 10 drops. All experiments were at $18.5^oC$ with kinesin concentration of 3.5 $\mu M$.

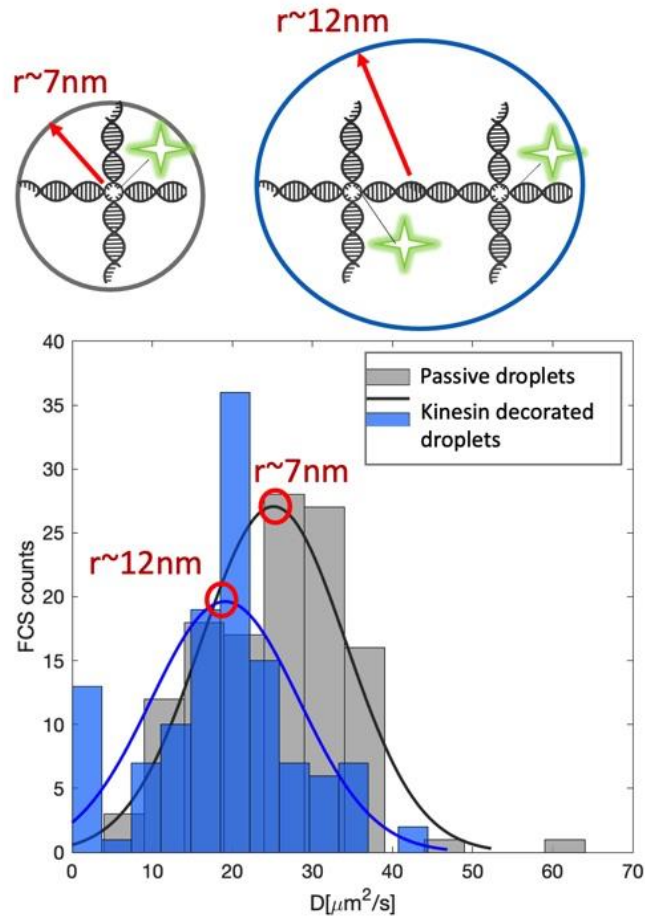

**Figure S5: Distribution of nanostar sizes in the dilute phase in passive and equilibrium samples. Distribution of molecules is obtained from** fluorescent correlation spectroscopy measurements of dilute phase for both kinesin decorated DNA droplets and passive droplets. - DNA-kinesin concentration is $3.5\ \mu M$ and $T = 20^o C$ .

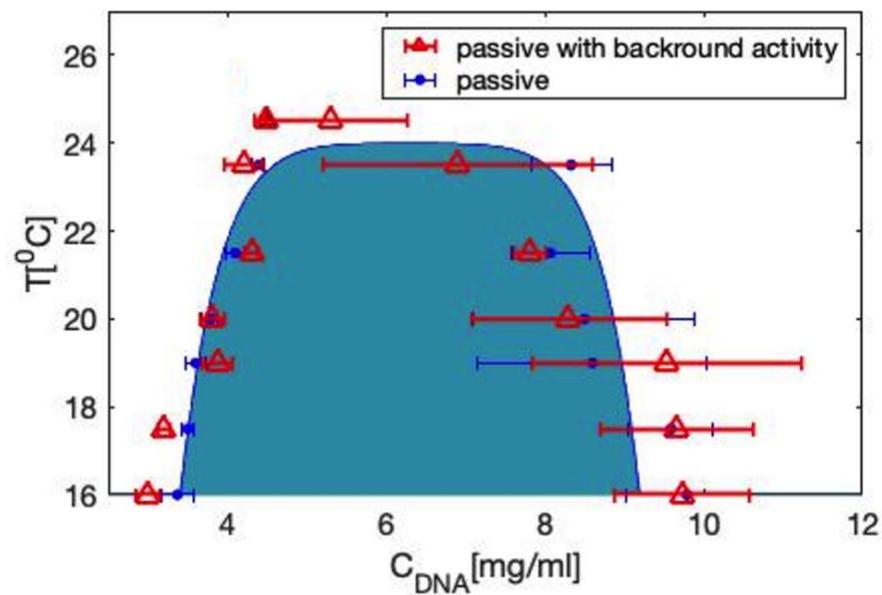

**Figure S6: Phase diagram of passive LLPS and active-LLPS lacking network-droplet coupling**. Active-LLPS was created by replacing DNA-kinesin with streptavidin based clusters which generated flow speeds of $0.6\ \mu m/s$. Such phase diagram is equivalent to the equilibrium LLPS.

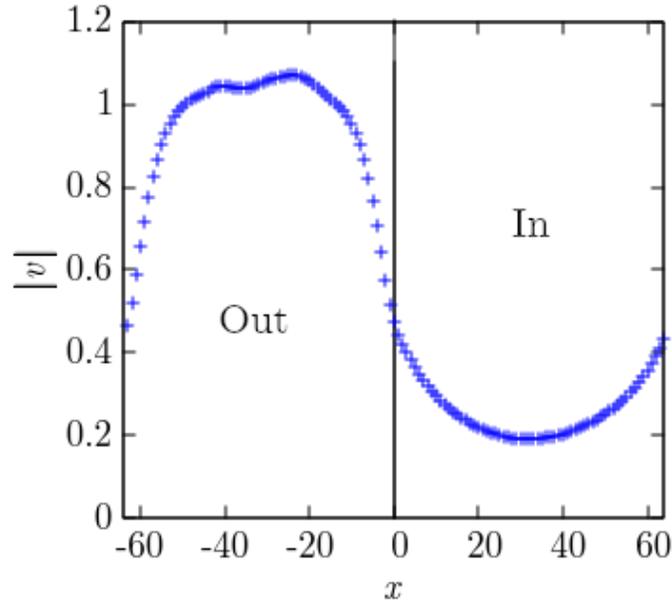

**Figure S7: Interfacial velocity from numerical simulations.** Average speed as a function of distance to the interface. Interface is located at $x = 0$, its width is $\Delta x \sim 3$. Positive $x$ values are in the droplet, and negative $x$ values are outside of the droplet in the active phase. This is done by simulating the model for $a = -1$, $\alpha = -10$, for an initially fully phase separated state. The velocity is averaged over the coordinate y (parallel to the interface), and time. Due to periodic boundary conditions, the speed profile decreases towards $x = -64$ and increases towards $x = 64$.

**Supplementary Videos**

**Video 1: Interaction of DNA nanostar droplets and active fluid**. **Part 1:** DNA droplet with no DNA-kinesin. Droplet is advected by MT flows using traditional streptavidin- biotin motor clusters. Overlayed video of DNA droplet (cyan), embedded in a MT active gel (red). **Part 2:** Dense droplets with no DNA-kinesin are advected and deformed by the reconfigurable MT network powered by DNA-kinesin nanostars in the dilute phase. The video is compatible to the data in Fig 1,c. Temperature is 19 $^oC$; DNA-kinesin concentration is 2.5 $\mu M$; scale bar, 50 $\mu m$.

**Video 2: LLPS in passive and active samples during temperature ramp-up. (Left)** Equilibrium LLPS containing DNA nanostars but no active fluid. **(right)** with DNA-kinesin in an active MT gel. Kinesin concentration 2.5 $\mu M$. Temperature indicated on the side panel. Each temperature step lasts 45 min. Scale bar, 100 $\mu m$.

**Video 3: Dynamics of LLPS with DNA-kinesin nanostars during ATP depletion.** DNA-kinesin depletes the available ATP leading to decrease in active flow speeds. During the slow down new DNA droplet form. To visualize formation of new droplets the contrast was enhanced, which saturated large droplets. Videos taken at $T = 19\ ^o$ C, DNA-kinesin concentration 3.5 $\mu M$.

**Video 4: Interactions of a droplet with an active fluid in absence of DNA-kinesin.** MT-Droplets interactions on droplet surface in a system lacking DNA-kinesin. Droplet is advected by MT flows using traditional streptavidin- biotin motor clusters. Projection layer of MTs on the surface of the DNA droplet is 11 $\mu m$ from the surface. **Part 1-3:** visualization of MT flows on 3 different droplet. Temperature is 17 $^oC$, active fluid speed is between $0.3 - 1\ \mu m/s$, scale noted by axis.

**Video 5: Interactions of a droplet with an active fluid in the presence of DNA-kinesin.** Visualization of MT-Droplets interactions in a DNA-kinesin droplet. Droplet is advected, deformed, and broken by MT flows powered by DNA-kinesin in the dilute phase. The droplets align with the long axis of the MTs bundles on the interface. **Part 1-3:** visualization of MT flows on 3 different droplet. Projection layer of MTs on the surface of the DNA droplet is 5.5 $\mu m$ from the surface. The videos are complimentary to the data in Fig 5,b. Temperature is $17.5^oC$, external fluid velocity is at $0.25\ \mu m/s$, kinesin concentration of 3.5 $\mu M$, scale bar noted by axis.

**Video 6: Interfacial velocity profile.** Two-dimensional confocal image of beads embedded in a DNA droplets. Droplets are faintly labeled with a YOYO dye. The 200 $\mu m$ beads are larger the nanostars core-to-core distances, indicating local deformations. Temperature is $17.5^oC$,

scale bar 100 $\mu m$. **Part 1:** DNA droplet with no kinesin-DNA. Droplet is advected by active flows, which are generated by streptavidin-kinesin clusters. External fluid velocity is 1 $\mu m/s$. **Part 2:** Droplets in a system with kinesin-DNA nanostars. Droplet is advected and deformed by active flows. Kinesin concentration, 3.5 $\mu M$.

**Video 7: Interfacial flow profile in numerical simulations.** Numerical integration of equations (1) of the main text inside and outside the coexistence region. We change activity $\alpha$ between the two runs to cross the coexistence region. Inside the coexistence region; **(Left)** Low activity, $\alpha = -10$, we observe the arrested phase separation described in the main text, in which a fully phase separated initial state breaks down until it reaches a steady state of very dynamic finite size droplets; **(Right)** high activity, $\alpha = -50$, active stresses are strong enough to stabilize the uniform state, and the system is stirred by the nematic fully mixing both species. The parameters are $a_Q = -a_\phi = b_Q = b_\phi = \gamma = \lambda = \eta = 1$ , $K = 5 \cdot 10^{-6}$ , $M = 0.1$, k=0.004. The integration is done on a square lattice of $128 \times 128$, with a lattice spacing of $\Delta x = 0.1$ and a time step $\Delta t = 0.001$.